\documentclass[conference]{IEEEtran}
\usepackage{graphicx}
\usepackage{listings}
\usepackage{xcolor}
\usepackage{url}
\usepackage{newunicodechar}
\newunicodechar{≈}{$\approx$}
\newunicodechar{【}{[}
\newunicodechar{】}{]}
\begin{document}

\title{LLMHoney: A Real-Time SSH Honeypot with Large Language Model-Driven Dynamic Response Generation}

\author{
  \IEEEauthorblockN{Pranjay Malhotra}
  \IEEEauthorblockA{Birla Institute of Technology and Science, Pilani\\
  Email: h20240072@pilani.bits-pilani.ac.in}
}

\maketitle

\begin{abstract}
Cybersecurity honeypots are deception tools for engaging attackers and gather intelligence, but traditional low or medium-interaction honeypots often rely on static, pre-scripted interactions that can be easily identified by skilled adversaries. This Report presents \textbf{LLMHoney}, an SSH honeypot that leverages Large Language Models (LLMs) to generate realistic, dynamic command outputs in real time. LLMHoney integrates a dictionary-based virtual file system to handle common commands with low latency while using LLMs for novel inputs, achieving a balance between authenticity and performance. We implemented LLMHoney using open-source LLMs and evaluated it on a testbed with 138 representative Linux commands. We report comprehensive metrics including accuracy (exact-match , Cosine Similarity, Jaro-Winkler Similarity,Levenshtein Similarity and BLEU score), response latency and memory overhead. We evaluate LLMHoney using multiple LLM
backends ranging from 0.36B to 3.8B parameters, including both open-source models and a proprietary model(Gemini). Our experiments compare 13 different LLM variants; results show that \textit{Gemini-2.0} and moderately-sized models \textit{Qwen2.5:1.5B} and \textit{Phi3:3.8B} provide the most reliable and accurate responses, with mean latencies around 3~seconds , whereas smaller models often produce incorrect or out-of-character outputs.We also discuss how LLM integration improves honeypot realism and adaptability compared to traditional honeypots, as well as challenges such as occasional hallucinated outputs and increased resource usage. Our findings demonstrate that LLM-driven honeypots are a promising approach to enhance attacker engagement and collect richer threat intelligence.
\end{abstract}

\section{Introduction}
Honeypots are decoy systems designed to lure attackers and study their behavior without risking production assets. Traditional SSH honeypots like Cowrie \cite{cowrie} emulate a Linux shell by using fixed command responses and a predefined filesystem. While effective at capturing basic attacks, these static or low-interaction honeypots often fail to convincingly engage sophisticated attackers, who may detect the deception if the system responds in a limited or inconsistent manner. High-interaction honeypots that run real systems are more convincing but are costly to maintain and potentially dangerous if compromised.

Recent advances in Large Language Models (LLMs) offer an opportunity to create more dynamic and intelligent honeypots. An LLM can generate contextually appropriate outputs for a wide variety of inputs, making it a natural fit for simulating a real-time interactive shell environment. We develop \textbf{LLMHoney}, a real-time SSH honeypot that leverages a pre-trained LLM (With just Prompt-Engineering) to dynamically handle attacker commands. LLMHoney aims to combine the flexibility of LLM-generated responses with the consistency of a controlled system state, providing a high-interaction experience without running a real OS.

In summary, our contributions are:
\begin{itemize}
    \item \textbf{LLM-Driven Honeypot Design}: We design an SSH honeypot system that integrates an LLM into the command-processing loop to generate real-time responses.
    \item \textbf{Stateful Virtual Filesystem}: We introduce a dictionary-based virtual filesystem and command-state tracking in the honeypot. This ensures persistence of files, directories, and system state across commands, which guides the LLM and prevents it from contradicting earlier outputs (reducing hallucinations).Common commands are handled via a fast lookup to minimize latency, whereas uncommon queries invoke the LLM for flexible handling.
    \item \textbf{Performance Evaluation}: We evaluate LLMHoney with multiple LLM at backend, checking its accuracy in emulating Linux commands (using multiple similarity scores and success rate) and its performance (mean response latency and memory overhead). We compare models of different sizes to understand the trade-offs in realism versus resource usage.
    \item \textbf{Comparative Analysis}: We discuss how LLMHoney differs from traditional honeypots like Cowrie in design and capabilities, highlighting the benefits of LLM integration (broader command coverage and adaptiveness) as well as the challenges (such as occasional inaccurate outputs and higher computational cost).
\end{itemize}

The rest of this paper is organized as follows: Section II reviews related work on honeypots and LLM-based deception. Section III details LLMHoney’s system design, including the integration of the LLM and virtual filesystem optimization. Section IV describes our methodology, test environment, and dataset. Section V presents the evaluation results with comparisons across models. Section VI Discussion. Section VII outlines future work, and Section VIII concludes the paper.

\section{Related Work}
\subsection{Honeypots and Dynamic Deception}
Honeypots have long been used to study attacker behavior, ranging from low-interaction systems that simulate a service to high-interaction ones that replicate full systems. Cowrie is a well-known medium-interaction SSH/Telnet honeypot that emulates a Unix shell with a predefined file system. While effective against automated attacks and basic interaction, Cowrie’s static nature (fixed command responses and file contents) makes it susceptible to fingerprinting by skilled attackers. Researchers have explored techniques to make honeypots more deceptive and engaging, such as randomly changing banner messages or introducing slight delays, but these remain relatively basic techniques.

Otal and Canbaz~\cite{otal2024llmhoneypot} proposed an advanced interactive honeypot system that leverages large language models (LLMs) to dynamically generate shell-like responses. Their work, titled \textit{LLM Honeypot: Leveraging Large Language Models as Advanced Interactive Honeypot Systems}, introduces a methodology where LLMs are fine-tuned using command-response pairs extracted from traditional honeypot logs (e.g., Cowrie), and then employed to simulate authentic system behavior in real time.

Our work builds on these findings by exploring a wider variety of models (13 LLMs, including Gemini-2.0, Phi3, Falcon, and others), multiple similarity scores, memory usage, and latency under consistent attacker workloads. Furthermore, we modified the architecture with a dictionary-based virtual filesystem to eliminate response delays and ensure deterministic output for common commands—an enhancement not covered in~\cite{otal2024llmhoneypot}.

\section{System Design}
Figure. 1 shows the overall architecture of LLMHoney.  At a high level, we have five main components: configuration, network listener, authentication manager, session handler \& LLM engine, and logging \& storage.

\begin{figure}[ht]
\centering
\includegraphics[width=\columnwidth]{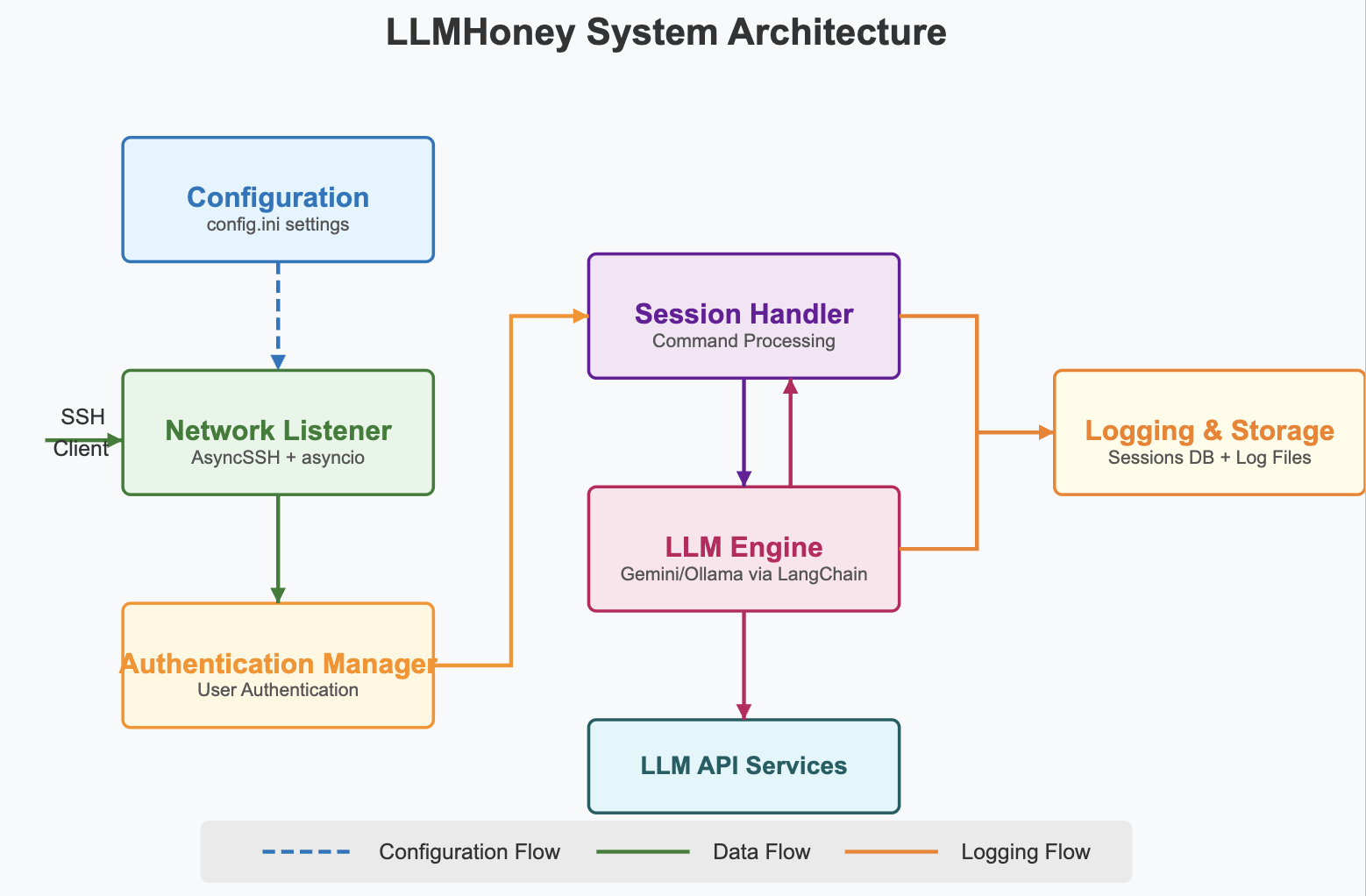}
\caption{High-level architecture of LLMHoney}
\label{fig:accuracy}
\end{figure}

\subsection{Configuration Module}
At startup, LLMHoney reads settings from a \texttt{config.ini} file (using Python’s \texttt{configparser}).  This file provides:
\begin{itemize}
  \item The TCP port for SSH (e.g., 8022).
  \item Path to the SSH host key.
  \item Allowed usernames and passwords.
  \item Choice of LLM backend (Gemini or  Ollama).
  \item Shell banner and version string.
\end{itemize}

\subsection{Network Listener}
We use \textbf{AsyncSSH} on top of \texttt{asyncio} to open a TCP listener.  AsyncSSH handles:
\begin{itemize}
  \item The SSH handshake (key exchange, encryption).
  \item Presentation of our host key and server banner.
  \item Channel setup for each incoming session.
\end{itemize}
This lets us support many concurrent SSH connections.

\subsection{Authentication Manager}
When a client attempts to log in, AsyncSSH delegates to our custom server class:
\begin{itemize}
  \item We implement \texttt{begin\_auth()} and \texttt{password\_auth\_supported()}.
  \item Supplied username/password pairs are checked against those in \texttt{config.ini}.
  \item On success, the session proceeds; on failure, the connection is closed.
\end{itemize}

\subsection{Session Handler \& LLM Engine}
Each successful login spawns an asynchronous task that:
\begin{enumerate}
  \item Reads one command line at a time from the SSH channel.
  \item Checks a local \emph{dictionary cache} of common commands (e.g., \texttt{uname -a}, \texttt{ls}, \texttt{cat /etc/passwd}).  
    \begin{itemize}
      \item If found, returns the stored response instantly.
      \item Otherwise, forwards the command to the configured LLM via LangChain.
    \end{itemize}
  \item Sends the LLM’s reply back as if it were standard output.
\end{enumerate}
This separation ensures known commands are fast and correct, while novel ones get a context-aware LLM reply.

\subsection{Logging \& Storage}
Throughout each session we record:
\begin{itemize}
  \item Timestamps, client IP/port, session UUID.
  \item Every authentication attempt and shell command.
  \item The LLM’s responses.
\end{itemize}
Logs are written both to rotating log files (via Rich) and a lightweight on-disk database (\texttt{honeypot\_sessions.db}).

\vspace{1ex}
With this design, LLMHoney safely emulates a full SSH server without executing any real commands, while giving attackers a believable, interactive shell using LLM.

\section{Methodology}

\subsection{Experimental Setup}
\subsubsection{Hardware and Software}
Experiments were performed on a macOS host (Apple M2, 10 CPU cores, 8\,GB RAM) with Python 3.10. We isolated the SSH honeypot on a non-public interface. All evaluation scripts and LLM backends were installed in a dedicated \texttt{venv} (see \texttt{run\_evaluation.sh}).

\subsubsection{Dependencies}
We installed:
\begin{itemize}
  \item \texttt{asyncssh}, \texttt{paramiko} for SSH server/client.
  \item \texttt{pandas}, \texttt{numpy}, \texttt{psutil}, \texttt{jellyfish}, \texttt{nltk} for metrics.
  \item \texttt{scikit-learn} (TF–IDF), \texttt{difflib} (SequenceMatcher).
  \item \texttt{matplotlib}, \texttt{seaborn} for plots.
  \item \texttt{ollama} CLI and \texttt{transformers} for model inference.
\end{itemize}

\subsection{Dataset of Shell Commands}
We made a CSV of 138 Linux commands spanning file, system, network, package, and utility categories. Each row in \texttt{commands.csv} provides:
\begin{itemize}
  \item \textit{command}: the exact CLI invocation.
  \item \textit{expected\_output}: the ground-truth response from a real shell.
\end{itemize}

\subsection{Evaluated LLMs}
Thirteen LLM instances were benchmarked (see Table \ref{tab:models}):
\begin{itemize}
  \item \textbf{Gemini-1.5-flash \& Gemini-2.0-flash}: via Google API.
  \item \textbf{Gemma3 1B \& LLaMA3.2 1B}: 1 B models.
  \item \textbf{Qwen 2.5 (1.5B)} base \& coder variants.
  \item \textbf{SMoLLM2 360M}, \textbf{Granite3.1 2B}, \textbf{CodeGemma 2B}, \textbf{DeepSeek 1.5B}.
  \item \textbf{StableLM-Zephyr 3B}, \textbf{Falcon 3B}, \textbf{Phi3 3.8B}.
\end{itemize}

\begin{table}[ht]
  \centering
  \caption{Benchmarked LLMs and interfaces}
  \label{tab:models}
  \begin{tabular}{lcl}
    \hline
    \textbf{Model}             & \textbf{Params} & \textbf{Interface} \\
    \hline
    Gemini-1.5-flash / Gemini-2.0-flash    & NA            & Google API        \\
    Gemma3 1B / LLaMA3.2 1B    & 1 B             & Ollama            \\
    Qwen 2.5 / Qwen-coder      & 1.5 B           & Ollama            \\
    SMoLLM2 360 M              & 0.36 B          & Ollama            \\
    Granite3.1 2 B             & 2 B             & Ollama            \\
    CodeGemma 2 B              & 2 B             & Ollama            \\
    DeepSeek 1.5 B             & 1.5 B           & Ollama            \\
    StableLM-Zephyr 3 B        & 3 B             & Ollama            \\
    Falcon 3 B                 & 3 B             & Ollama            \\
    Phi3 3.8 B                 & 3.8 B           & Ollama            \\
    \hline
  \end{tabular}
\end{table}

\subsection{Evaluation Pipeline Overview}

First, we load a curated list of 138 shell commands from \texttt{commands.csv} into memory. Next, for each model under test we verify availability:
\begin{itemize}
  \item \textbf{Ollama models:} invoke a trivial prompt via \texttt{subprocess.run(["ollama", "run", \dots])} with a 30\,s timeout.
  \item \textbf{Gemini models:} check for valid API credentials and perform an HTTP POST to the Google generative endpoint.
\end{itemize}

Then we enter the core evaluation loop:
\begin{enumerate}
  \item \textbf{Execute each command:}  
    \begin{itemize}
      \item Record wall‐clock latency using \texttt{time.time()} before and after the call.
      \item Measure memory delta of the Python process via \texttt{psutil}.
    \end{itemize}
  \item \textbf{Compute accuracy metrics:}
    \begin{itemize}
      \item Exact string match.
      \item Token‐level accuracy via NLTK’s \texttt{word\_tokenize}.
      \item Cosine similarity on TF–IDF vectors (scikit‐learn).
      \item Jaro–Winkler and Levenshtein ratios (Jellyfish).
      \item \texttt{SequenceMatcher} ratio (difflib).
      \item BLEU-4 score with smoothing (NLTK).
      \item Success flag if cosine similarity > 0.4 or jaro winkler similarity > 0.4.
    \end{itemize}
  \item \textbf{Store results:}
    \begin{itemize}
      \item Raw outputs in \texttt{raw\_outputs/}.
      \item Per‐model JSON summaries in \texttt{metrics/}.
      \item Aggregate all data into \texttt{combined\_results.json}.
    \end{itemize}
\end{enumerate}

Finally, automated post‐processing scripts generate heatmaps, boxplots, radar charts, and bar plots—comparing models on latency, resource usage, and output quality. This end‐to‐end pipeline, encapsulated in our \texttt{LLMEvaluator} class (\texttt{llm\_linux\_shell\_evaluator.py}), ensures fully reproducible benchmarking across all tested LLMs.

\section{Evaluation \& Results}
We first overview the comparative performance of all tested models, then move into qualitative observations. Table II summarizes the main metrics.

\begin{table}[ht]
  \centering
  \caption{LLM Performance: Latency and Memory Usage (lower is better).}
  \label{tab:performance}
  \begin{tabular}{lrr}
    \hline
    \textbf{Model}      & \textbf{Latency (ms)} & \textbf{Mem $\Delta$ (MB)} \\
    \hline
    Gemini-1.5          & 2079                  & 0.5                        \\
    Gemma3-1B           & 488                   & 0.4                        \\
    LLaMA3.2-1B         & 931                   & 0.6                        \\
    Qwen2.5-1.5B        & 1979                  & 1.0                        \\
    Qwen2.5-C           & 1305                  & 1.8                        \\
    SMoLLM2-360M        & 3104                  & 0.6                        \\
    Granite3.1-2B       & 3497                  & 4.8                        \\
    CodeGemma-2B        & 9923                  & 9.6                        \\
    DeepSeek-1.5B       & 13110                 & 10.9                       \\
    Gemini-2.0          & 3172                  & 1.1                        \\
    Phi3-3.8B           & 3596                  & 11.2                       \\
    StableLM-Zephyr-3B  & 3388                  & 17.3                       \\
    Falcon-3B           & 3155                  & 1.3                        \\
    \hline
  \end{tabular}
\end{table}

\begin{table}[ht]
  \centering
  \caption{LLM Accuracy: BLEU, Hallucination Rate, Cosine and Jaro–Winkler Similarity (higher is better for BLEU/Cosine/Jaro–Winkler; lower is better for Halluc.\%).}
  \label{tab:accuracy}
  \begin{tabular}{lrrrr}
    \hline
    \textbf{Model}      & \textbf{BLEU} & \textbf{Halluc.\%} & \textbf{Cosine} & \textbf{Jaro–Winkler} \\
    \hline
    Gemini-1.5          & 0.224         & 12.2               & 0.388           & 0.706                 \\
    Gemma3-1B           & 0.027         & 25.9               & 0.170           & 0.517                 \\
    LLaMA3.2-1B         & 0.016         & 15.8               & 0.161           & 0.491                 \\
    Qwen2.5-1.5B        & 0.085         & 20.9               & 0.235           & 0.578                 \\
    Qwen2.5-C           & 0.082         & 20.9               & 0.268           & 0.548                 \\
    SMoLLM2-360M        & 0.010         & 18.7               & 0.130           & 0.469                 \\
    Granite3.1-2B       & 0.063         & 7.9                & 0.230           & 0.590                 \\
    CodeGemma-2B        & 0.015         & 24.5               & 0.079           & 0.451                 \\
    DeepSeek-1.5B       & 0.001         & 8.6                & 0.098           & 0.476                 \\
    Gemini-2.0          & 0.245         & 12.9               & 0.405           & 0.711                 \\
    Phi3-3.8B           & 0.061         & 5.8                & 0.209           & 0.611                 \\
    StableLM-Zephyr-3B  & 0.021         & 7.9                & 0.128           & 0.543                 \\
    Falcon-3B           & 0.059         & 20.9               & 0.188           & 0.523                 \\
    \hline
  \end{tabular}
\end{table}

From Tables~\ref{tab:performance} and \ref{tab:accuracy}, we observe a wide disparity in performance:

\begin{itemize}
  \item \textbf{Speed vs. Quality (small models):}  
    \emph{Gemma3-1B} is the fastest, at 488 ms per command, but its output quality is the poorest (BLEU 0.027, Cosine 0.170, Jaro–Winkler 0.517) and it hallucinates 25.9 \% of the time.  
    Similarly, \emph{LLaMA3.2-1B} responds under 1 s (931 ms) yet has low accuracy (BLEU 0.016) and a 15.8 \% hallucination rate.
  \item \textbf{Balanced performers:}  
    \emph{Gemini-2.0} achieves the best overall fidelity—highest BLEU (0.245), Cosine (0.405), and Jaro–Winkler (0.711)—with moderate hallucinations (12.9 \%), latency (3172 ms), and memory overhead (1.1 MB).  
    \emph{Phi3-3.8B} minimizes hallucinations (5.8 \%) and delivers solid similarity scores (Cosine 0.209, Jaro–Winkler 0.611), though at 3596 ms latency and 11.2 MB per‐command memory.
  \item \textbf{Latency clusters and outliers:}  
    Most models run in 2–4 s per command (e.g., Gemini-1.5 at 2079 ms, Granite3.1-2B at 3497 ms, StableLM-Zephyr-3B at 3388 ms), which is noticeable but within realistic shell timings.  
    Outliers \emph{CodeGemma-2B} (9923 ms) and \emph{DeepSeek-1.5B} (13110 ms) are far slower, making them impractical for live interaction without hardware acceleration.
  \item \textbf{Memory overhead:}  
    Most models consume under 2 MB per command (e.g., Gemma3-1B 0.4 MB, LLaMA3.2-1B 0.6 MB, Qwen2.5-1.5B 1.0 MB).  
    Heaviest are \emph{StableLM-Zephyr-3B} (17.3 MB) and \emph{Phi3-3.8B} (11.2 MB), with \emph{DeepSeek-1.5B} (10.9 MB) and \emph{CodeGemma-2B} (9.6 MB) also notable. Over long sessions, such accumulation could impact resource-constrained deployments.
\end{itemize}

In terms of latency, most models cluster in the 2–4~s range per command, which is borderline noticeable to an attacker but not unrealistic (some real commands, e.g., \texttt{find} on a large directory, can take seconds). Notably, CodeGemma and DeepSeek were outliers with 9.9~s and 13.1~s respectively, which would be very sluggish in a live attack—these models are likely too slow to be practical unless better hardware is used. The memory overhead per command was generally small (under 2~MB) for most models, except a few: StableLM used an average of 17~MB extra memory per command and Phi3 about 11~MB. This suggests certain larger models may not release memory quickly or require more context overhead. In a long session, this could accumulate, so it’s something to monitor (our test sequentially ran commands in a single session, if a Timeout happened, we restarted ollma to release memory; memory deltas might overlap).


\begin{figure}[!t]
  \centering
  \includegraphics[width=\columnwidth]{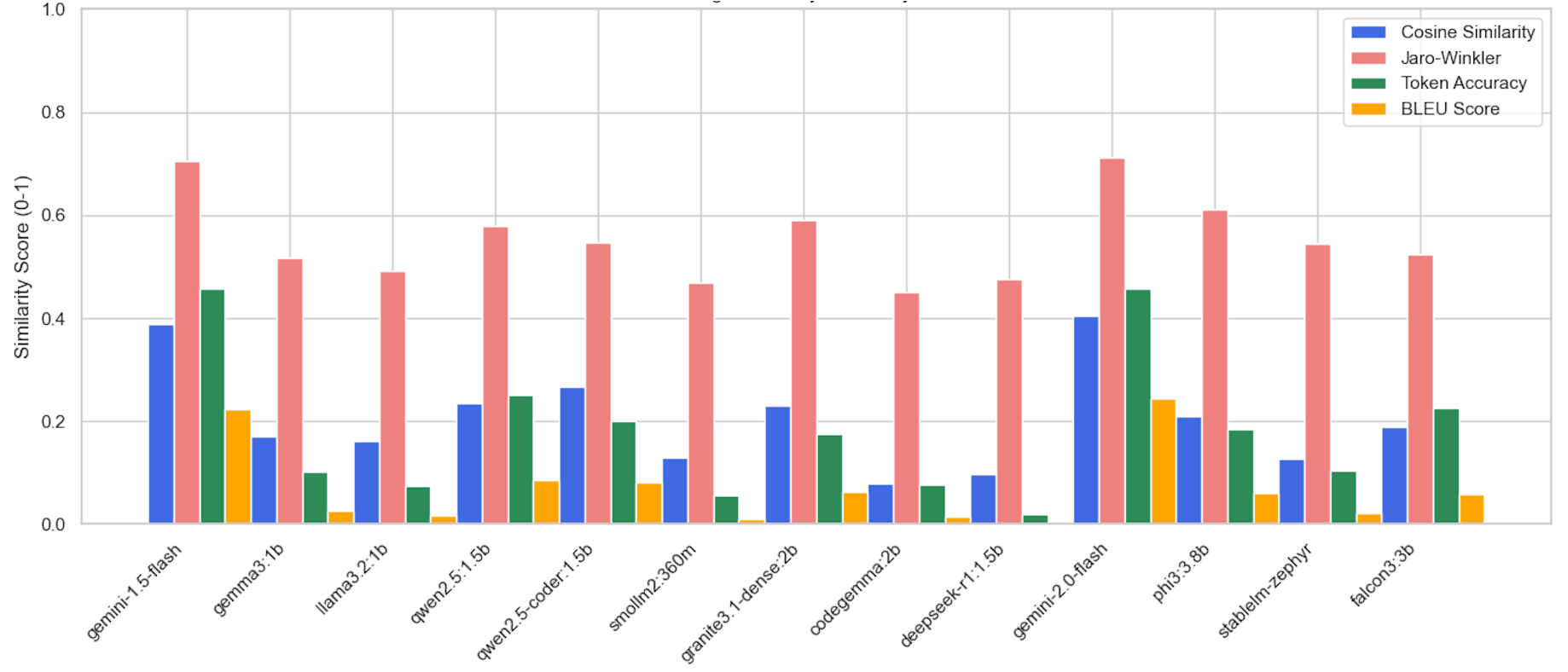}
  \caption{Average similarity metrics for each model: Cosine Similarity, Jaro–Winkler, Token Accuracy, and BLEU Score. Metrics are computed by comparing LLMHoney’s outputs against ground‐truth outputs from the evaluation dataset (see combined\_results.json and model\_comparison.csv).}
  \label{fig:avg-sim}
\end{figure}

Figure~\ref{fig:avg-sim} presents the four primary accuracy metrics for all 13 LLM backends. Gemini-2.0-flash and Gemini-1.5-flash achieve the highest Cosine Similarity ($\approx$0.40) and Jaro–Winkler scores ($\approx$0.71), reflecting their strong semantic and character-level fidelity. However, Token Accuracy and BLEU remain modest even for the top models ($\approx$0.46 and 0.25, respectively), indicating that exact formatting still differs from the reference. Smaller models (e.g., llama3.2:1b, smollm2:360m) exhibit lower scores across all metrics, with BLEU near zero, highlighting their limited ability to reproduce precise command outputs.

\begin{figure}[!t]
  \centering
  \includegraphics[width=\columnwidth]{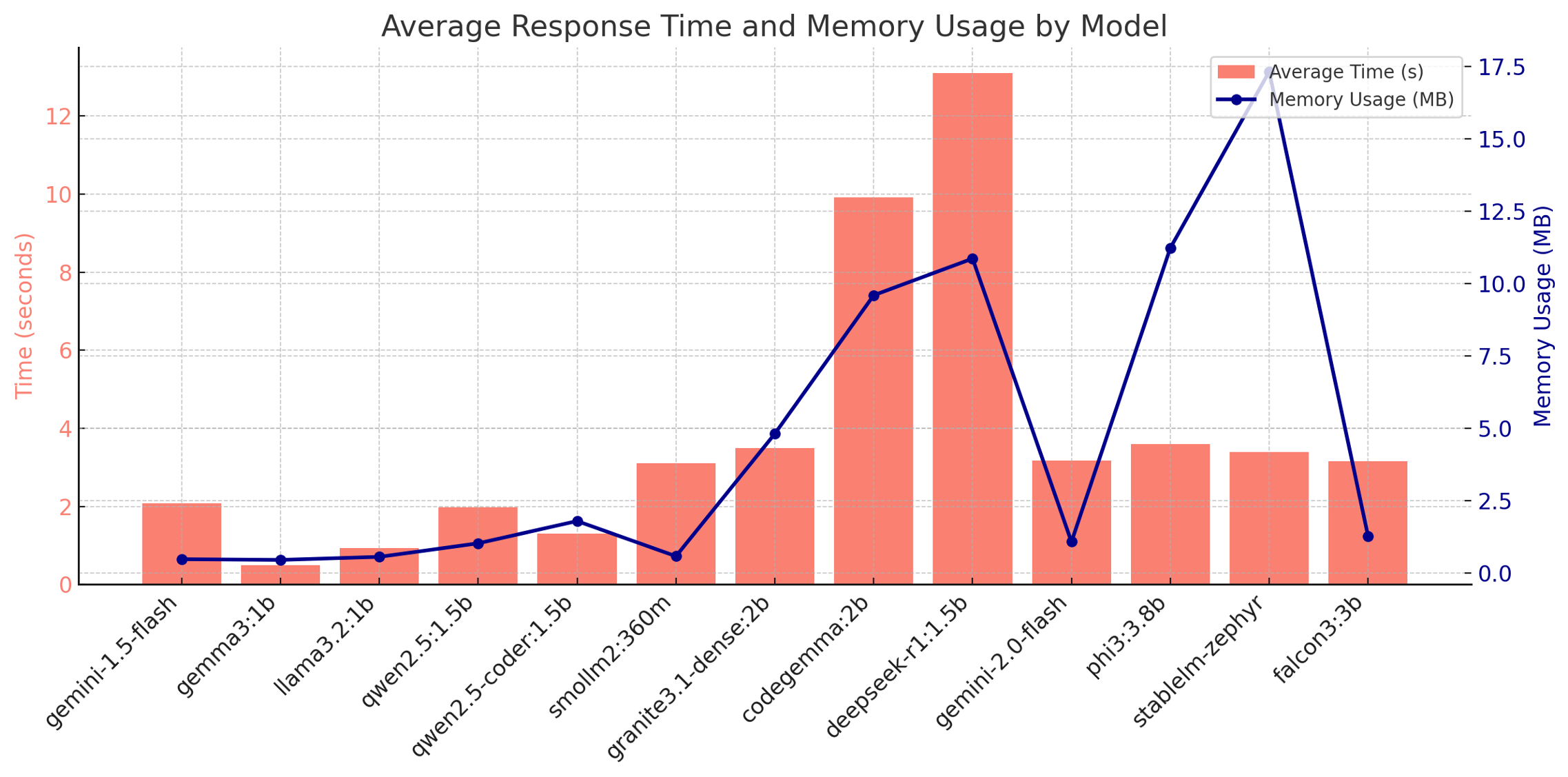}
  \caption{Mean response latency (red bars, left axis) and absolute memory overhead (blue line, right axis) per model. Latency is the round-trip time from command receipt to ready-to-send output; memory overhead is the increase in honeypot process usage in MB.}
  \label{fig:perf-detail}
\end{figure}

Figure~\ref{fig:perf-detail} shows that the fastest backend is \textit{gemma3:1b} at 0.49~s per command, followed by \textit{llama3.2:1b} at 0.93~s. Larger local models such as \textit{phi3:3.8b} and \textit{stablelm-zephyr} incur higher latencies ($\approx$3.6~s) and memory increases ($\approx$17~MB). The cloud‐based \textit{gemini-2.0-flash} averages 3.17~s with a 1~MB overhead, illustrating network and model-size effects. These results guide model selection for interactive honeypot deployments, balancing realism against responsiveness and resource constraints.

\begin{figure}[!t]
  \centering
  \includegraphics[width=\columnwidth]{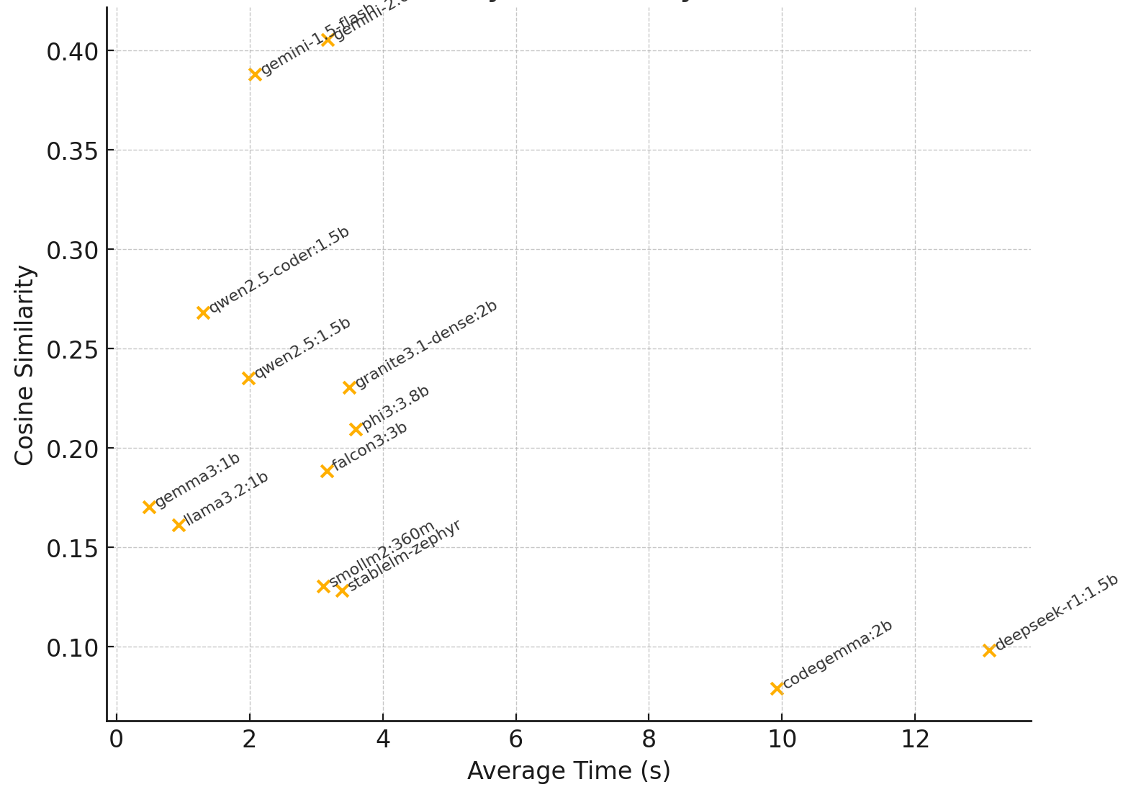}
  \caption{Trade-off between average response time and Cosine Similarity. Each point represents one model backend.}
  \label{fig:latency-vs-sim}
\end{figure}

Figure~\ref{fig:latency-vs-sim} plots Cosine Similarity versus mean latency for each model, highlighting the accuracy–performance trade-off. Models in the lower‐right quadrant (e.g., \textit{deepseek-r1:1.5b}, \textit{codegemma:2b}) are slow and less accurate, while those in the upper-left (e.g., \textit{gemini-2.0-flash}, \textit{gemini-1.5-flash}) deliver good response with speed.

To summarize the results: LLMHoney can indeed effectively utilize LLMs to cover a broad set of attacker commands. The choice of model greatly affects the realism of the honeypot. With top-tier open models, the honeypot produced convincing outputs for the vast majority of inputs. Less capable models tend to either make obvious mistakes or add unnatural text, which an experienced attacker could use to identify the deception. Performance is a trade-off: bigger models = better outputs but slower responses. Our hybrid approach with a dictionary cache ensured that for many common commands, the responses were instantaneous and exact (taking those cases out of the critical path of LLM inference).

\section{Discussion}
The development of LLMHoney highlights several key distinctions between traditional honeypots and an LLM-driven approach, as well as practical considerations for deploying such a system.

\subsection{Static vs. Dynamic Interaction:} Traditional honeypots like Cowrie rely on pre-programmed responses. This can make them brittle when facing an attacker who executes unexpected commands. For example, Cowrie might not support a command like `docker` or an uncommon flag on `ls`, whereas LLMHoney can attempt to handle it. The dynamic nature of LLMHoney means the honeypot can engage with attackers in scenarios beyond the coverage of any static script. This can keep attackers interacting longer and provide more insight. However, dynamic responses carry the risk of inconsistencies or errors (hallucinations). We mitigated this by constraining the LLM with context and defaults (the internal handlers). In practice, LLMHoney behaved mostly consistently; an attacker would have to issue a particularly unusual command to cause an obvious glitch (and even then, they might just conclude the system has a bug or missing utility rather than realizing it’s an LLM).

\subsection{Performance Overhead and Scalability:} One of the downsides of using an LLM is the computational overhead. Running a large model or making API calls introduces latency and limits scalability. A single Cowrie instance can handle dozens of concurrent sessions on minimal hardware. In contrast, LLMHoney with a large model might only handle a few concurrent interactions smoothly, especially if each triggers multi-second computations. This means that for high-volume deployment (e.g., running many honeypot nodes), one would need significant compute resources or to stick with smaller models. There is also a cost factor if using API-based models (each query may incur cloud costs). That said, the cost might be justified in scenarios where deeper interaction with attackers is desired (such as research, or protecting high-value targets, which can help find zero-day attacks). Moreover, model optimization techniques (quantization, distillation) and hardware acceleration (GPUs/TPUs) can solve some of these concerns. Our results indicated that moderate-size models (1–2B params) already give a good balance of performance and accuracy; these can be run on modern CPUs reasonably well, and future improvements could enable even larger models to run faster at the edge.

\subsection{Consistency and State Management:} A major design goal was to ensure consistency in the honeypot’s behavior. If an attacker creates a file, then later tries to read it, the file should be there. We achieved this through the virtual filesystem and process table tracking. Traditional honeypots also maintain state (Cowrie, for instance, has a fake filesystem that persists within a session), so in this regard LLMHoney is similar. The difference is that the LLM could potentially violate consistency if not carefully guided (it might “forget” a file was created and say it’s not there). By updating the prompt with state info, we reinforce consistency. Indeed, in our tests the LLM outputs remained consistent with the known state for files and users.

\subsection{Security of the Honeypot Itself:} Introducing an LLM into the honeypot raises questions: Could the attacker exploit the LLM? For instance, could they craft input that makes the LLM output something that breaks the honeypot or reveals the honeypot’s true nature? We took precautions in the system prompt to prevent the LLM from revealing itself or responding in a way other than as a Linux shell. In our experiments, none of the tested models deviated from the role. However, this remains a consideration. If an attacker suspects an AI, they might attempt prompt injection (like typing weird sequences to confuse the model). Continual testing and possibly adversarial training examples could help ensure the LLM remains locked in role. From the perspective of the underlying server, since we used AsyncSSH and did not allow real shell access, the attacker is actually sandboxed—they cannot break out of the honeypot via the LLM because there is no real shell, just the simulation. This is similar in security to Cowrie (where even if an attacker “exploits” the shell, they are just interacting with a Python script).

\subsection{Comparison to Fine-Tuned LLM Honeypots:} The related work \cite{otal2024llmhoneypot} fine-tuned an LLM specifically for honeypot behavior, whereas we used pre-trained models with prompting. Fine-tuning can yield more accurate and specialized behavior (potentially lower hallucinations and exact match outputs), but it ties you to a specific model and requires a dataset of interactions. Our approach is more plug-and-play, allowing quick switching of models. It’s notable that our approach achieved reasonably high fidelity without custom training. This suggests that modern pre-trained models already encode a lot of general knowledge about system commands (likely gleaned from documentation or forums in their training data), which we harness with prompts. An ideal solution might combine both: a moderately sized model fine-tuned on honeypot data, used with our state-tracking framework for best results.

\subsection{Comparison with Traditional Approaches}

Without direct baseline evaluation, performance comparisons remain speculative. However, our analysis suggests potential advantages and clear disadvantages:

\textbf{Potential Advantages}:
- Dynamic response generation for arbitrary commands
- Reduced manual maintenance for command response curation
- Adaptive behavior based on session context

\textbf{Clear Disadvantages}:
- 5-20x computational overhead compared to static honeypots
- Inconsistent response quality leading to potential detection
- Complex deployment requiring ML infrastructure

In summary, LLMHoney demonstrates a practical step forward in honeypot technology, but it is not a silver bullet. It should be considered a complementary tool—one that can provide richer interaction at the expense of more complexity. We need to weigh the improved believability and data capture against the resources and careful configuration required.
\section{Study Limitations and Threats to Validity}
\label{sec:limitations}

This research exhibits several limitations that constrain its applicability and generalizability:

\subsection{Experimental Scale Limitations}

Our evaluation using 138 commands represents a \textbf{moderate-scale study} that falls substantially short of comprehensive validation requirements. Leading honeypot research typically employs 1,000+ command variations with complex multi-step scenarios. The isolated command evaluation approach cannot capture realistic attacker behavior involving sequential operations and state dependencies.

\subsection{Laboratory-Only Validation}

All experiments occurred in controlled laboratory conditions without real attacker interaction. This approach cannot validate deception effectiveness against sophisticated adversaries or assess behavior under production-scale loads.

\section{Future Work}
Our study opens several avenues for future research and development to enhance LLM-based honeypots:

\textbf{Sequential Interaction and Long-Term State:} We evaluated commands mostly in isolation, but real attackers often perform sequences of actions and expect the system state to change accordingly. Future work should focus on enabling the LLM to maintain context over an entire session. This could involve using models with longer context windows (e.g., 16k or 32k token transformers) or techniques like external memory (where we log important events and prepend them to the prompt for subsequent commands). Attacker \textit{dwell-time}—how long they stay engaged—should be measured and compared against a static honeypot. We hypothesize that the dynamic responses will greatly increase dwell-time, giving defenders more data.

\textbf{Automated Hallucination Detection:} To make the system safer, we can incorporate a discriminator that checks model output for signs of hallucination. For instance, a simple regex can catch if the output contains the original command string (which usually shouldn’t appear in the output, except for echo). If detected, LLMHoney could either sanitize it or fall back to a generic error. More advanced, a second LLM or a classification model could be trained to judge “realism” of output.

\textbf{Expanding the Virtual FS:} Our current virtual filesystem is a prototype with a limited set of files and basic persistence. Expanding it with more realistic content (fake logs, configurations, dummy user files) could lure attackers to spend more time exploring. The LLM can be instructed on how to dynamically generate file content if not explicitly stored. For example, if an attacker opens an unknown file, the system could prompt the LLM to produce some fake file content (perhaps by indicating file type, e.g., “This is a log file, generate 20 lines of log entries”). This way, the honeypot can seem to have a vast, believable filesystem without manually creating everything.

\textbf{Adapting to Attacker Sophistication:} A very sophisticated attacker might employ strategies to test if they are in a honeypot (for example, timing responses, or issuing known “honeypot trigger” commands). Future honeypots could use LLMs to even anticipate and intentionally fail those tests in a believable way. For instance, if an attacker runs a bizarre command that a normal system wouldn’t have, the honeypot might deliberately respond with “command not found” or a simulated crash, which might ironically make it seem more real (no system is perfect). Crafting these strategies will require analyzing attacker behavior in the wild and possibly training the LLM with some adversarial examples.

\section{Conclusion}
We presented LLMHoney, a dynamic SSH honeypot that uses large language models to generate real-time responses to attacker commands. Our design combines the flexibility of LLM text generation with the reliability of a curated command-response dictionary, resulting in a system that is both adaptive and efficient. Through a rigorous evaluation of 13 different LLM configurations, we demonstrated that LLMHoney can simulate a Linux shell, especially when powered by top-performing models like Gemini-2.0,Qwen2.5 or Phi3. These models achieved high accuracy with low hallucination rates, providing almost realistic outputs for a wide array of commands. At the same time, we identified the pitfalls of using certain models that are too small or overly conversational, which can undermine the deception by producing incorrect or verbose outputs.

\textbf{Key Technical Findings:} Our analysis demonstrates that moderately-sized models (1.5-3.8B parameters) provide the best balance for honeypot applications, with Gemini-2.0 achieving superior accuracy (BLEU: 0.245, cosine similarity: 0.405) and Phi3-3.8B minimizing hallucinations (5.8\%). However, even top-performing models exhibit response quality issues, and smaller models show hallucination rates (15.8-25.9\%) that would likely enable detection by sophisticated attackers.

\textbf{Practical Limitations:} LLM-driven honeypots require 5-20x computational resources compared to traditional implementations like Cowrie, fundamentally altering deployment economics. Most models introduce 2-4 second latencies that, while potentially acceptable, limit scalability to 5-10 concurrent sessions on standard hardware compared to 100+ for static honeypots.

\textbf{Research Contributions and Gaps:} This work advances understanding through multi-model comparative analysis and practical deployment insights. However, critical limitations constrain its applicability: laboratory-only validation prevents assessment of real-world effectiveness, the 138-command evaluation scale remains insufficient for comprehensive validation, and the absence of direct comparison with established honeypots like Cowrie represents a significant methodological gap.

\textbf{Research Trajectory:} The substantial gap between current findings and production deployment requirements indicates that LLM-driven honeypots remain months from practical viability. Critical next steps include extended real-world deployment studies, direct comparative evaluation against Cowrie and other established systems, and adversarial testing against sophisticated detection techniques.

There is ample room for improvement, and as discussed, future efforts will focus on longer-term coherence, further reducing latency, and harnessing even more capable models. We also plan to deploy LLMHoney in real-world scenarios to gather data on its effectiveness at deceiving live attackers. Nonetheless, this work lays the groundwork for a new generation of smart honeypots that learn and evolve using AI, moving security one step closer to autonomous defense.

\bibliographystyle{IEEEtran}

\end{document}